\documentclass[twocolumn,showpacs,preprintnumbers,amsmath,amssymb,pre]{revtex4}
\usepackage{graphics}
\usepackage{epsfig}

\begin{document}

\title{Perturbation method to model enamel caries progress}

\author{Katarzyna D. Lewandowska}
 \email{kale@amg.gda.pl}
 \affiliation{Department of Physics and Biophysics, \\
	Medical University of Gda\'nsk, \\
	ul. D\c{e}binki 1, 80-211 Gda\'nsk, Poland.}

\author{Tadeusz Koszto{\l}owicz}
 \email{tkoszt@pu.kielce.pl}
 \affiliation{Institute of Physics, \'Swi\c{e}tokrzyska Academy,\\
         ul. \'Swi\c{e}tokrzyska 15, 25-406 Kielce, Poland.}

\date{\today}


\begin{abstract}
We develop a theoretical model of the carious lesion progress caused by acids diffusing into the tooth enamel from the dental plaque. The acids react with static hydroxyapatite, which leads to demineralization of the enamel, and consequently to the development of the carious lesion. The model utilizes the diffusion-reaction equations with one static and one mobile reactant where the reaction term is proportional to the product of concentrations of acids and of mineral. The changes of concentrations are calculated approximately by means of a perturbation method. The analytical approximate solutions are compared with the numerical ones and experimental data.
\end{abstract}

\maketitle

\section{Introduction}

The carious lesion progress in dental enamel has been extensively studied experimentally 
~\cite{abqde,ac,ae,ale,bl,bqade,ca,ca1,cfh,daeg,db,dbd,f,fdc,fdc1,fm,fr,ga,gea,gray1,gray2,hg,hghp,hmpbh,itra,kpssm,mz,mzlkm,tddg,whff,zkm}. The experiments were performed {\it in vitro} with extracted teeth being demineralized in
buffer solutions of organic acids at different values of {\it pH}~\cite{abqde,ac,ae,ale,bl,bqade,ca,ca1,cfh,daeg,db,dbd,f,fdc,fdc1,fm,fr,ga,gea,gray1,gray2,hg,hghp,hmpbh,itra,kpssm,mz,mzlkm,tddg,whff,zkm}. These studies examined the time evolution of the depth of the carious lesion~\cite{ca1,fdc,fm,f,fdc1,fr,ga,gea,hg}, the mineral loss in the dental
enamel~\cite{gray2,hg,whff,hmpbh,mzlkm}, the rate of enamel
dissolution~\cite{abqde,bqade,gray1,gray2,hmpbh,hghp,zkm}, the affect of various
factors on the progress of caries such as concentrations and
{\it pH} of buffer or inhibitors~\cite{gray1,gray2,whff}, the
concentration profiles of hydroxyapatite ({\it HA}) (which is the main
component of enamel)~\cite{ac,itra,bl,cfh,ale,ae,abqde,ga,mzlkm,zkm,kpssm,tddg,gea,daeg}. However, there are only a few attempts at theoretical description of the carious lesion progress. 

The formation of carious lesion of enamel starts when concentration
of organic acids in the dental plaque reaches the sufficient value and {\it pH} of
dental plaque lowers below the appropriate point. Then, the organic
acids diffuse in undissociated and/or dissociated form inward the enamel and react with the mineral to form soluble calcium and phosphate ions (or complexes)
\cite{hg,fdc,mz,fdc1}. Thus, theoretical model should describe the diffusion of acids
inside the enamel followed by the reaction with static hydroxyapatite. The diffusion-reaction equations for the system with one static and one mobile reactants are usually chosen as~\cite{ah,bs}
	\begin{eqnarray}
\label{1} \frac{{\partial A(x,t)}}{\partial t}&=&D\frac{\partial^2 A(x,t)}{\partial x^2}-R(x,t),\\
\label{2} \frac{\partial C(x,t)}{\partial t}&=&-R(x,t) ,
	\end{eqnarray}
where $A$ is the concentration of diffusing particles, $D$ - their diffusion coefficient, $C$ is the concentration of static reactant, and $R$ denotes the reaction term, which is chosen within the mean field approximation as~\cite{ah,bs}
	\begin{equation}\label{3}
R(x,t)=kA^{\rm m} (x,t)C^{\rm n} (x,t) ,
	\end{equation}
$k$ is the reaction rate; the parameters $n$ and $m$ (which may be non-integer) are determined experimentally. The explicit solutions of Eqs.~(\ref{1}) and~(\ref{2}) with the reaction term [Eq.~(\ref{3})] was found only for very few special cases~\cite{murray,ah}. Thus, one is usually interested in finding some general characteristic functions of the diffusion-reaction system such as the time evolution of reaction front, the width of depletion zone etc.~\cite{gr}, which are usually found by means of numerical methods of solving the diffusion-reaction equations and computer simulations. To obtain approximate concentration profiles $A$ and $C$ simplifying methods are used, as for example the quasistatic~\cite{koza} or perturbation~\cite{thktw,tkhkw,tykhw} ones. 

To theoretically describe the caries one often adopts assumptions which oversimplify the problem. Some authors consider only one of Eqs.~(\ref{1}) and~(\ref{2}). Additionally the reaction term $R$ is taken in oversimplified form, where only one reactant is taken into account. For example, only Eq.~(\ref{1}) was considered in~\cite{whff} with $R=k(C_{\rm s}-C)$, where $C_{\rm s}$ is a constant related to `solubility' of the solute. In~\cite{r} the reaction term was chosen in the form $R(x,t)=k(x,t)A(x,t)$. The model based on Eq.~(\ref{2}) only was studied in~\cite{gray2} with $R(x,t)=kC^{\rm n}$, and in~\cite{bqade} with $R(x,t)=k(C_{\rm s}-C)$. Both Eqs.~(\ref{1}) and~(\ref{2}) were considered in~\cite{se,db}, where $R(x,t)=kA$. We add that there were also considered the models based only on the diffusion equation without chemical reactions included through the `effective'
diffusion coefficient~\cite{hg,hghp,hmpbh,cfh}. Very special diffusive theoretical models were used to describe some caries characteristics such as temporal evolution of the amount of hydroxyapatite in the buffer solution released form the enamel during its dissolution~\cite{gray2,gray1,se,db,ca}, time evolution of the caries limit~\cite{ca1,fdc,fdc1,hg,f,fm,kl}, and concentration profiles of fluoride~\cite{cfh} and of hydrogen ions~\cite{r} inside the enamel in the stationary state.
As far as we know, there the theoretical concentrations of the mineral as a function of time have not yet been obtained. 

Characteristic feature of carious lesion process is the creation of the surface layer in which the loss of hydroxyapatite is relatively small. However, an appearance of this layer complicates theoretical models because the acid particles that diffuse through the layer do not chemically react with hydroxyapatie. The existence of a surface layer is not taken into account explicitly in the theoretical models mentioned above. 

In our paper we use both of Eqs.~(\ref{1}) and~(\ref{2}) to describe the carious lesion process. The surface layer is also included in our considerations. We focus our attention on finding theoretical formulas of the concentration profiles of both reactants during the caries progress. To 
 solve Eqs.~(\ref{1}) and~(\ref{2}) we use a perturbation method. The analytical approximate solutions are compared with the numerical ones of the diffusion--reaction equations and experimental data.

\section{Carious lesion process}

The enamel consists of hydroxyapatite crystals $Ca_{10}(PO_4)(OH)_2$ in $92-94\%$ by weight~\cite{zero}. These crystals are organized in larger forms called prisms. The intercrystalline and interprismatic spaces of enamel are filled with
water~\cite{mz, zero}. Because of spaces between crystals and prisms, enamel is a microporous material~\cite{zero,fdc,mz,mzlkm,ca1,ae,ale}. Besides {\it HA} the enamel includes inorganic factors, mostly fluoride and carbonate. In addition more then $40$ trace elements can occur in the tooth mineral. Organic matrix represents less then $1\%$ by weight of the enamel~\cite{zero}. The surface of dental enamel is covered by the dental plaque which mainly consists of microorganisms, saliva, leftovers and mucus. Oral microorganisms metabolize simple sugars coming from diet~\cite{mz,fdc} to the organic acids (e.g., acetic or
lactic). As we have mentioned previously, the formation of carious lesion of enamel starts when
concentration of organic acids in dental plaque reaches sufficient
value and {\it pH} of the dental plaque lowers below the appropriate point.
Then the organic acids diffuse into the enamel~\cite{f,fdc,hg,hmpbh,bqade}. The acids
can be transported in dissociated or undissociated form that
depends on {\it pH} of the dental plaque~\cite{f,fdc,ca1,hg,gray2}. After achieving the enamel interior, acid reacts with the mineral according to the chemical formula~\cite{gray1,hg,fdc,mz,fdc1,hmpbh}
\begin{displaymath}
 Ca_{10}(PO_4)_6(OH)_2 + 8H^+ \rightarrow 10Ca^{2+} + 6HPO^{2-}_4 + 2H_{2}0,
\end{displaymath}
where the phosphate ions have an acidic form determined by the {\it pH} of the system. The products of reaction are inert for the caries progress~\cite{f}.
\begin{figure}
		\includegraphics{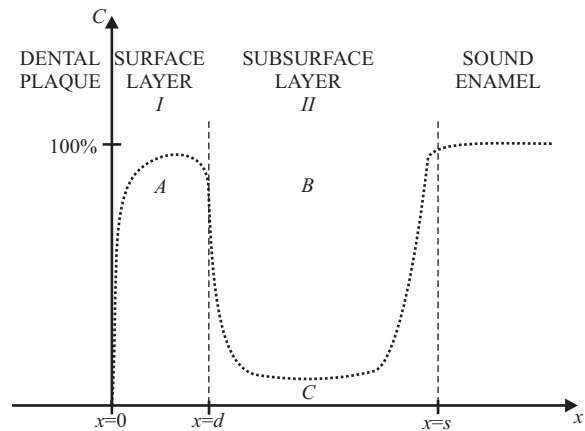}
	\caption{\label{fig:fig1}The schematic view of the tooth enamel. The doted line
represents the concentration of hydroxyapatite.}	
\end{figure}
It is commonly accepted that the products of the reaction, calcium ions and
phosphate ions (or complexes), are transported out of the
enamel by the diffusion~\cite{mz,ca1,f,fdc,hg,fr,ale,abqde,hmpbh,bqade}. There are two distinguishing stages in the formation
of caries. In the first stage an apparently intact surface layer is created~\cite{mz,fdc1,f,fm,fr,mzlkm,bqade} (see Fig.~\ref{fig:fig1}). This is
the layer where the loss of mineral is small in comparison to the
content of the mineral in the sound enamel. Two possible mechanisms
responsible for creating the surface layer have been proposed.
First, the appearance of protective agents prevents acids from
dissolving the enamel~\cite{fdc,fdc1,f,fm,hg,fr,ca1}. Second, the surface layer
appears as a result of the combination of the dissolution and
reprecipitation processes~\cite{mz,mzlkm}. The thickness of this layer after
reaching a maximum value remains unchanged later on~\cite{f,fdc1}. Dissolution of
the subsurface mineral (situated below the apparently intact surface
layer) occurs in the second stage of the formation of carious
lesion. The dissolution reaction takes place in a restricted region
of the enamel called the reaction zone (the inner zone or the decalcification region). At the beginning the zone is
placed right below the outer enamel surface but after exhausting the
hydroxyapatite in that region the reaction zone moves inside the
enamel~\cite{f,fdc,hg}. Host factors involved in the caries process are as follows: composition and structure of the enamel \cite{fdc,hmpbh}, {\it pH} and buffer concentration \cite{hg,fr,mzlkm}, kind of acid \cite{fr,mzlkm,hmpbh}, mineral content gradient \cite{ae,ale}, composition of saliva, age of tooth, environment, diet, hygiene, etc.

\section{Model}

The system under study is three-dimensional, but it is assumed to be homogeneous in the plane perpendicular to the $x$ axis, which is normal to the tooth surface. So, it can be treated as one-dimensional. The system described in the previous section is presented schematically
in Fig.~\ref{fig:fig1}. This scheme is based on the qualitative descriptions of caries presented in the papers~\cite{ac,tddg} and on the plots of experimental concentration profiles given in~\cite{ac,itra,bl,cfh,ale,ae,abqde,ga,mzlkm,zkm,kpssm,tddg,gea,daeg}. The border between the surface layer and the subsurface one is
not sharp in realistic system, but for simplicity we assume that
these regions are separated by the plane located at $x=d$.

In the following we label the surface layer that does not change in time as $I$ and the subsurface layer as $II$. Let us denote the concentration of hydrogen
ions as $A$ in the $I$ region, $B$ in the $II$ region and the
concentration of {\it HA} in the $II$ region as $C$. We assume that the
pure diffusion occurs in the region $I$, and there is
the diffusion with chemical reactions in the region $II$. The ions diffuse in both regions with diffusion coefficient $D$. There arises a problem with a choice of the reaction term, as the parameters $m$ and $n$ occurring in Eq.~(\ref{3}) have not been unambiguously determined experimentally or theoretically for the caries process. Beside we note that $m$ and $n$ depend on the chemical composition and structure of the tooth enamel, which is a personality trait, and on tooth environment, which seem to be uncontrollable in real systems. Here we assume that the reaction term is given by Eq.~(\ref{3}) with $m=n=1$. This assumption simplifies the procedure of solving Eqs.~(\ref{1}) and~(\ref{2}). We presume that the case with $m>1$ and/or $n>1$ will not change significantly the qualitative form of the solutions (see for example~\cite{bs}, where the cases of $m=n=1$ and $m=n=2$ are considered). Thus, we choose the equations describing the caries as follows

	\begin{eqnarray}
\label{4} \frac{\partial A(x,t)}{\partial t}&=&D\frac{\partial^2
A(x,t)}{\partial x^2} ,\\
\label{5} \frac{\partial B(x,t)}{\partial t}&=&D\frac{\partial^2
B(x,t)}{\partial x^2}-kB(x,t)C(x,t) ,\\
\label{6} \frac{\partial C(x,t)}{\partial t}&=&-kB(x,t)C(x,t) .
	\end{eqnarray}

At the initial time $t=0$ we assume that the enamel is free of
acid and {\it HA} forms a homogeneous medium of the
concentration $C_0$. This assumption provides the initial
conditions
	\begin{equation}\label{7a}
A(x,0)=B(x,0)=0, 
	\end{equation}
	\begin{equation}\label{7}
C(x,0)=C_0.
	\end{equation}
The dental plaque is the reservoir of the acid, so we assume that
at the border of the enamel ($x=0$) the concentration of
the acid $A_0$ is constant. The thickness of the enamel is very large compared to the region where the acid concentration varies, so we treat the enamel as semi-infinite medium. It is rather
obvious that the acid concentration and flux are continuous at the
border between the $I$ and $II$ regions. Thus we adopt the following
boundary conditions	
	\begin{equation}\label{8a}
A(0,t)=A_0, A(d,t)=B(d,t) ,
	\end{equation}
	\begin{equation}\label{8b}
\left.\frac{\partial A(x,t)}{\partial x}\right|_{\rm x=d}
=\left.\frac{\partial B(x,t)}{\partial x}\right|_{\rm x=d} ,
	\end{equation}
	\begin{equation}\label{8}
B(\infty ,t)=0 .
	\end{equation}
Because there is no exact analytic method to solve Eqs.~(\ref{4})-(\ref{6}), we use the perturbation technique to find approximate solutions.

\section{Perturbation method}

To use the perturbation method we first transform Eqs.~(\ref{4})-(\ref{8}) to the dimensionless form using the substitutions:
	\begin{equation}\label{9}
x=\rho x_{\rm s} ,\ t=\tau t_{\rm s} ,
	\end{equation}
where $\rho$ and $\tau$ denote the dimensionless position and time, $x_{\rm s}$ and $t_{\rm s}$ are constants of the dimension of space and time, respectively. Using Eq.~(\ref{9}) we obtain the transport equations
	\begin{eqnarray}
\label{10} \frac{\partial a(\rho,\tau)}{\partial \tau}&=&\frac{\partial^2
a(\rho,\tau)}{\partial \rho^2} ,\\
\label{11} \frac{\partial b(\rho,\tau)}{\partial \tau}&=&\frac{\partial^2
b(\rho,\tau)}{\partial \rho^2}-\varepsilon b(\rho,\tau)c(\rho,\tau) ,\\
\label{12} \frac{\partial c(\rho,\tau)}{\partial \tau}&=&-\varepsilon r b(\rho,\tau)c(\rho,\tau) ,
	\end{eqnarray}
initial conditions
	\begin{equation}\label{13}
a(\rho,0)=b(\rho,0)=0, c(\rho,0)=1,
	\end{equation}
and boundary conditions
	\begin{equation}\label{14a}
a(0,\tau)=1, a(\rho_{\rm d},\tau)=b(\rho_{\rm d},\tau) ,
	\end{equation}
	\begin{equation}\label{14b} 
\frac{\partial
a(\rho,\tau)}{\partial \rho}|_{\rho=\rho_{\rm d}}=\frac{\partial
b(\rho,\tau)}{\partial \rho}|_{\rho=\rho_{\rm d}} ,
	\end{equation}
	\begin{equation}\label{14} 
b(\infty ,\tau)=0,
	\end{equation}
where 
	\begin{eqnarray}
a(\rho,\tau)&\equiv&\frac{A(\rho x_{\rm s},\tau t_{\rm s})}{A_0},\nonumber\\
b(\rho,\tau)&\equiv&\frac{B(\rho x_{\rm s},\tau t_{\rm s})}{A_0},\nonumber\\ 
c(\rho,\tau)&\equiv&\frac{C(\rho x_{\rm s},\tau t_{\rm s})}{C_0},\label{c}
	\end{eqnarray}
	\begin{equation}
\label{podst} r=\frac{A_0}{C_0}, \qquad \rho_{\rm d}=\frac{d}{x_{\rm s}},
	\end{equation}
and	
	\begin{equation}\label{15}
\varepsilon =kC_0t_{\rm s} .
	\end{equation}
Equations~(\ref{10}) and~(\ref{11}) are derived under condition that the parameters $x_{\rm s}$ and $t_{\rm s}$ fulfill the relation
	\begin{equation}\label{16}
x_{\rm s}=\sqrt{Dt_{\rm s}} .
	\end{equation}

Let us assume now that the parameter $\varepsilon$ is small enough
($\varepsilon\ll 1)$) to apply the perturbation method (with respect
to this parameter) to find the concentration of {\it HA} as a function of
space and time.

Within the perturbation method the concentrations are given as
	\begin{eqnarray}
\label{17} a(\rho,\tau)&=&\sum^\infty_{\rm n=0}a_{\rm n}(\rho,\tau)\varepsilon^{\rm n} ,\\
\label{18} b(\rho,\tau)&=&\sum^\infty_{\rm n=0}b_{\rm n}(\rho,\tau)\varepsilon^{\rm n} ,\\
\label{19} c(\rho,\tau)&=&\sum^\infty_{\rm n=0}c_{\rm n}(\rho,\tau)\varepsilon^{\rm n} .
	\end{eqnarray}
Substituting Eqs.~(\ref{17})-(\ref{19}) to Eqs.~(\ref{10})-(\ref{14}) and comparing the functions of the same order with respect to the parameter $\varepsilon$ occurring at both sides of these equations, we get the equations of the zeroth order
	\begin{eqnarray}
\label{20} \frac{\partial a_0(\rho,\tau)}{\partial \tau}&=&\frac{\partial^2
a_0(\rho,\tau)}{\partial \rho^2},\\
\label{21} \frac{\partial b_0(\rho,\tau)}{\partial \tau}&=&\frac{\partial^2
b_0(\rho,\tau)}{\partial \rho^2},\\
\label{22} \frac{\partial c_0(\rho,\tau)}{\partial \tau}&=&0,
	\end{eqnarray}
with the initial conditions
	\begin{eqnarray}
\label{ic1a} a_0(\rho,0)&=&0 ,\\
\label{ic2a} b_0(\rho,0)&=&0 ,\\
\label{ic3a} c_0(\rho,0)&=&1, 
	\end{eqnarray}
and boundary ones
	\begin{eqnarray}
\label{bc1a} a_0(0,\tau)&=&1 ,\\
\label{bc2a} a_0(\rho_{\rm d},\tau)&=&b_0(\rho_{\rm d},\tau), \\
\label{bc3a} \frac{\partial a_0(\rho,\tau)}{\partial
\rho}|_{\rho=\rho_{\rm d}}&=&\frac{\partial b_0(\rho,\tau)}{\partial
\rho}|_{\rho=\rho_{\rm d}} .
	\end{eqnarray}
For the order of $n=1,2,3,\ldots$ we get the following equations
	\begin{eqnarray}
\label{23}\frac{\partial a_{\rm n}(\rho,\tau)}{\partial \tau}&=&\frac{\partial^2
a_{\rm n}(\rho,\tau)}{\partial \rho^2},\\
\label{24}\frac{\partial b_{\rm n}(\rho,\tau)}{\partial \tau}&=&\frac{\partial^2
b_{\rm n}(\rho,\tau)}{\partial \rho^2}\nonumber \\ 
& & -\sum^{\rm n-1}_{\rm k=0}
b_{\rm k}(\rho,\tau)c_{\rm n-k-1}(\rho,\tau),\\
\label{25}\frac{\partial c_{\rm n}(\rho,\tau)}{\partial \tau}&=&-r\sum^{\rm n-1}_{\rm k=0}
b_{\rm k}(\rho,\tau)c_{\rm n-k-1}(\rho,\tau),
	\end{eqnarray}
with the initial conditions
	\begin{eqnarray}
\label{ic1b} a_{\rm n}(\rho,0)&=&0 ,\\
\label{ic2b} b_{\rm n}(\rho,0)&=&0 ,\\
\label{ic3b} c_{\rm n}(\rho,0)&=&0 ,  
	\end{eqnarray}
and the boundary conditions
	\begin{eqnarray}
\label{bc1b} a_{\rm n}(0,\tau)&=&0 ,\\ 
\label{bc2b} a_{\rm n}(\rho_{\rm d},\tau)&=&b_{\rm n}(\rho_{\rm d},\tau), \\
\label{bc3b} \left.\frac{\partial a_{\rm n}(\rho,\tau)}{\partial
\rho}\right|_{\rm \rho=\rho_d}&=&\left.\frac{\partial b_{\rm n}(\rho,\tau)}{\partial
\rho}\right|_{\rm \rho=\rho_d} .
	\end{eqnarray}

We solve the above equations by means of the Laplace transform method \cite{cj}. Equations~(\ref{23})-(\ref{25}) become more and more difficult to solve when the order of perturbation method grows. In practice, it is possible to obtain explicit solutions up to the first order. In the next section we find the exact solutions for $a_{\rm i}$, $b_{\rm i}$ and $c_{\rm i}$ where $i=0,1$.

\section{Analytical and numerical results}

Solving Eqs.~(\ref{20})-(\ref{22}) with initial [Eqs.~(\ref{ic1a})-(\ref{ic3a})]
and boundary [Eqs.~(\ref{bc1a})-(\ref{bc3a})] conditions we obtain
	\begin{equation}\label{zeros}
a_0(\rho,\tau)=b_0(\rho,\tau)=
{\rm erfc}  \left[\frac{\rho}{2\sqrt{\tau}}\right],\quad c_0(\rho,\tau)=1,
	\end{equation}
where $
{\rm erfc}(u)=\frac{2}{\sqrt{\pi}}\int^\infty_u\exp(-\eta^2)d\eta$ is the
complementary error function. So, the zeroth order solutions take form of the pure diffusion without chemical reactions.

Equations of first order are
	\begin{eqnarray}\label{firste}
\nonumber\frac{\partial a_1(\rho,\tau)}{\partial
\tau}&=&\frac{\partial^2
a_1(\rho,\tau)}{\partial \rho^2}, \\
\frac{\partial b_1(\rho,\tau)}{\partial \tau}&=&\frac{\partial^2
b_1(\rho,\tau)}{\partial \rho^2}-b_0(\rho,\tau)c_0(\rho,\tau) ,\\
\nonumber\frac{\partial c_1(\rho,\tau)}{\partial
\tau}&=&-rb_0(\rho,\tau)c_0(\rho,\tau).
	\end{eqnarray}
When supplemented by the initial conditions $a_1(\rho,0)=b_1(\rho,0)=c_1(\rho,0)=0$ and the boundary ones $a_1(0,\tau)=0, a_1(\rho_{\rm d},\tau)=b_1(\rho_{\rm d},\tau)$ and $\frac{\partial a_1(\rho,\tau)}{\partial\rho}|_{\rho=\rho_{\rm d}}=\frac{\partial b_1(\rho,\tau)}{\partial
\rho}_{\rho=\rho_{\rm d}}$, the solutions of Eqs.~(\ref{firste}) are
	\begin{eqnarray}\label{firsts1}
\lefteqn{a_1(\rho,\tau)=}\nonumber\\
& & \frac{1}{4}\left[\left(\tau+\frac{(2\rho_{\rm d}+\rho)^2}{2}\right){\rm erfc}\left(\frac{2\rho_{\rm d}+\rho}{2\sqrt{\tau}}\right)\right.\nonumber\\
& & -(2\rho_{\rm d}+\rho)\sqrt{\frac{\tau}{\pi}}e^{-(2\rho_{\rm d}+\rho)^2/4\tau}\\
& &-\Big(\tau+\frac{(2\rho_{\rm d}-\rho)^2}{2}\Big){\rm erfc} \left(\frac{2\rho_{\rm d}-\rho}{2\sqrt{\tau}}\right)\nonumber\\
& &\left.+(2\rho_{\rm d}-\rho)\sqrt{\frac{\tau}{\pi}}e^{-(2\rho_{\rm d}-\rho)^2/4\tau}\right] ,\nonumber
	\end{eqnarray}
	\begin{eqnarray}\label{firsts2}
\lefteqn{b_1(\rho,\tau)=}\nonumber\\
& &\frac{1}{4}\left[\left(\tau+\frac{(2\rho_{\rm d}+\rho)^2}{2}\right){\rm erfc}\left(\frac{2\rho_{\rm d}+\rho}{2\sqrt{\tau}}\right)\right.\nonumber\\
& &-(2\rho_{\rm d}+\rho)\sqrt{\frac{\tau}{\pi}}e^{-(2\rho_{\rm d}+\rho)^2/4\tau}\\
& &-\Big(\tau+\frac{(4\rho_{\rm d}-3\rho)\rho}{2}\Big){\rm erfc}\left(\frac{\rho}{2\sqrt{\tau}}\right)\nonumber\\
& &\left.+(4\rho_{\rm d}-3\rho)\sqrt{\frac{\tau}{\pi}}e^{-\rho^2/4\tau}\right] ,\nonumber
	\end{eqnarray}
	\begin{equation}\label{firsts3}
c_1(\rho,\tau)=-r\left[\left(\tau+\frac{\rho^2}{2}\right) {\rm
erfc}\left[\frac{\rho}{2\sqrt{\tau}}\right]-\rho\sqrt{\frac{\tau}{\pi}}
e^{-\rho^2/4\tau}\right] .
	\end{equation}

\begin{figure}[h]
		\includegraphics[scale=0.85]{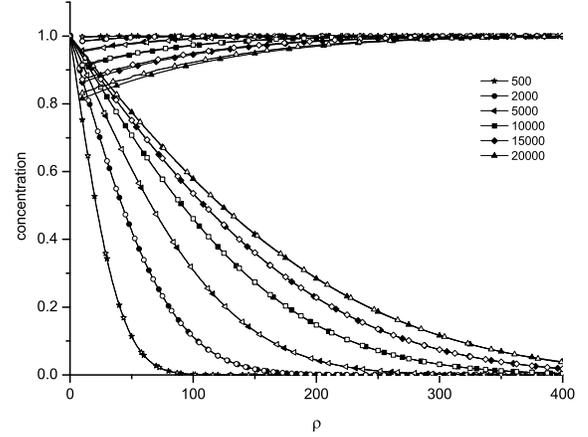}	
	\caption{\label{fig:fig.2}The concentrations of hydroxyapatite (increasing functions) and acid (decreasing functions) in the subsurface layer as a function of distance from the tooth surface for different times given in the legend, all quantities are given in dimensionless units. The parameters are $\varepsilon =10^{-5}$ and $r=1$. The continuous lines with black symbols represent the approximate solutions (\ref{firsta})--(\ref{firstc}), the continuous lines with white symbols--the numerical solutions of Eqs. (\ref{10})-(\ref{12}).}
	\end{figure}

\begin{figure}[h]
		\includegraphics[scale=0.85]{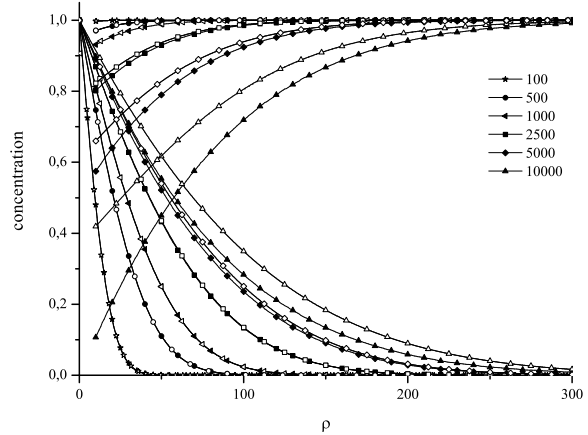}
	\caption{\label{fig:fig.3}The numerical and approximate analytical solutions as in Fig.~\ref{fig:fig.2} but for $\varepsilon =10^{-4}$ and $r=1$.}
\end{figure}

\begin{figure}[h]
		\includegraphics[scale=0.85]{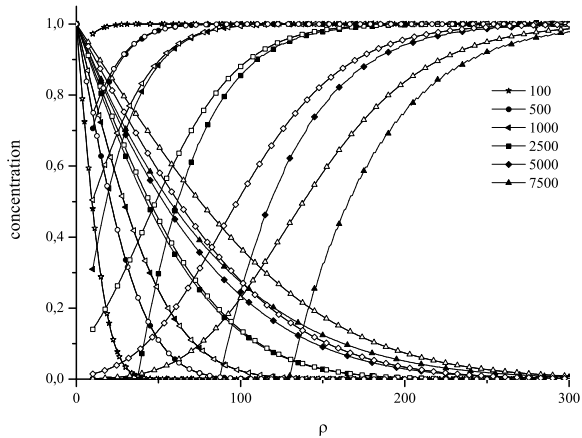}
	\caption{\label{fig:fig.4}The description is the same as in Fig. \ref{fig:fig.2} but for $\varepsilon =10^{-4}$ and $r=10$.}
\end{figure}

In Figs.~\ref{fig:fig.2}-\ref{fig:fig.4} we present the plots of the numerical solutions of Eqs. (\ref{10})-(\ref{12}) and the dimensionless concentrations of acid 
	\begin{equation}\label{firsta}
a(\rho,\tau)=a_0(\rho,\tau)+\varepsilon a_1(\rho,\tau),	
	\end{equation}
for $0<\rho<\rho_{\rm d}$,
	\begin{equation}\label{firstb}
b(\rho,\tau)=b_0(\rho,\tau)+\varepsilon b_1(\rho,\tau),
	\end{equation}
for $\rho > \rho_{\rm d}$ (the decreasing functions of $\rho$) and the dimensionless concentration of {\it HA} (the growing functions of $\rho$)
	\begin{equation}\label{firstc}
c(\rho,\tau)=c_0(\rho,\tau)+\varepsilon c_1(\rho,\tau) .
	\end{equation}
To use the numerical procedure we take the following approximations of the derivatives
$\partial c(\rho,\tau)/\partial \tau=[c(\rho,\tau)-c(\rho,\tau-\Delta\tau)]/\Delta\tau$, $\partial^2 c(\rho,\tau)/\partial\rho^2=[c(\rho+\Delta\rho,\tau)+c(\rho-\Delta\rho,\tau)-2c(\rho,\tau)]/(\Delta\rho)^2$, calculations were performed for $\Delta\rho=0.25$ and $\Delta\tau=0.01$.
The numerical solutions and approximate analytical ones are computed for few values of $\varepsilon$ and $r$ and for several values of time indicated in the legends of the plot; in all cases $\rho_{\rm d} =5$.

For $\varepsilon=10^{-5}$ and $r=1$ the numerical solutions (continuous lines with white symbols) and approximate analytical ones (continuous lines with black symbols) representing the acid concentrations coincide with each other (Fig.~\ref{fig:fig.2}). Such a coincidence occurs for larger values of the parameters only for relatively small times (Figs.~\ref{fig:fig.3} and~\ref{fig:fig.4}). Similar behavior is observed for the solutions representing the concentration of {\it HA}, but the difference between numerical and approximate analytical solutions are larger for the latter functions. Let us note that up to the first order approximation [Eqs.~(\ref{firsta})-(\ref{firstc})] the concentrations of diffusive reactant depend explicitly on parameter $\varepsilon$ whereas the concentrations of static reactant depend on the product of the parameters $\varepsilon r$. When $\varepsilon r$ is relatively small, then the loss of mineral is noticeable for relatively large times. For example, the $5\%$ loss of mineral is observed  after $t\approx6000$ if $\varepsilon r=10^{-5}$, after $t\approx1000$ if $\varepsilon r=10^{-4}$, and after $t\approx100$ if $\varepsilon r=10^{-3}$. In the plots presented in Fig.~\ref{fig:fig.2}-\ref{fig:fig.4} we assume that $\varepsilon r\ll 1$. Similar plots can be obtained for relatively large $\varepsilon r$ for sufficiently small times. Of course, the above remarks concern the dimensionless quantities. To express the functions in the dimensional variables one needs to set the values of $x_{\rm s}$ and $t_{\rm s}$ obtained on the basis of experimental data. 

\section{Comparison with experimental data}

The plots of {\it HA} shown in Fig.~\ref{fig:fig.2}-\ref{fig:fig.4} qualitatively agree with the experimental data. Fig.~\ref{fig:fig.2} is similar to Fig.~1 from \cite{mzlkm}, Fig.~1 from \cite{zkm} and Figs.~3~A and~3~B from~\cite{bqade}, Fig.~\ref{fig:fig.3} is like Fig.~3 from \cite{cfh},
Fig.~\ref{fig:fig.4} is in qualitative accordance with Fig.~1 from \cite{abqde}. 
Quantitative comparison of the model predictions with the experimental data appears to be difficult. Experimental results obtained at very similar conditions are often very different from each other. The loss of {\it HA} achieves $\sim80\%$ of the initial value within $14$ days according to \cite{zkm} but only $\sim40\%$ of the initial value within $89$ days as reported in the paper \cite{mzlkm}. This can be caused by individual features of each tooth. The amount of experimental data is too small to determine the influence of the various factors.

\begin{figure}
		\includegraphics[scale=0.8]{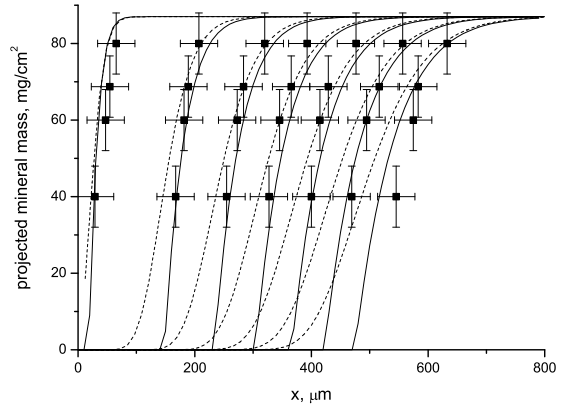}
	\caption{\label{fig:fig5}Theoretical concentration of {\it HA} ({\it continuous lines}) calculated for $t_{\rm f}=5.5$ h and $\Delta t=48\;{\rm h}$ and numerical solutions $C(x,t)$ ({\it dashed lines}), separated points represent experimental data taken from \cite{gea}.}
\end{figure}

\begin{figure}
		\includegraphics[scale=0.8]{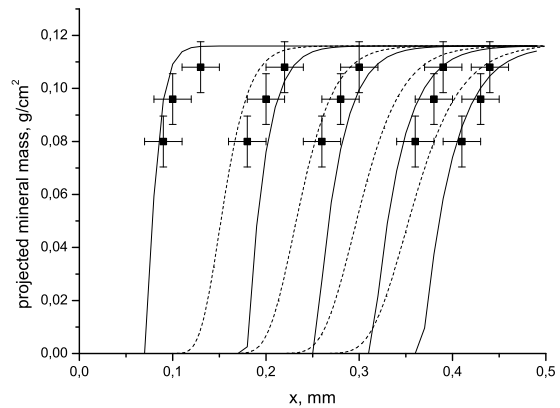}
	\caption{\label{fig:fig6}Experimental results and theoretical ones obtained for $t_{\rm f}=17$ h and $\Delta t=54$ h; additional description is in the text.}
\end{figure}

Despite the difficulties, we analyze the data taken from Fig.~1 presented in \cite{gea} and from Fig.~2 presented in \cite{daeg} ($C$ is given in special units ${\rm g/cm^2}$  commonly used in scanning microradiography). The experimental concentration profiles extracted from these figures and theoretical functions are shown in Figs.~\ref{fig:fig5} and~\ref{fig:fig6}, respectively. Taking into account how the experimental points are scattered, and including the accuracy of our readout, we estimated the error to be $32\;{\rm \mu m}$ for $x$ and $8\; {\rm mg/cm^2}$ for $C(x,t)$ in Fig.~\ref{fig:fig5} and $0.02\;{\rm mm}$ for $x$, $0.0096\;{\rm g/cm^2}$ for $C(x,t)$ in Fig.~\ref{fig:fig6}. These errors are marked in Figs.~\ref{fig:fig5} and \ref{fig:fig6}. The theoretical concentration profile expressed through dimensional variables is obtained from Eqs.~(\ref{podst})-(\ref{16}), (\ref{zeros}), (\ref{firsts3}), (\ref{firstc}), and it reads as
	\begin{eqnarray}\label{stwym}
C(x,t)&=&C_0-A_0C_0k\left[\left(t+\frac{x^2}{2D}\right){\rm erfc}\left(\frac{x}{2\sqrt{Dt}}\right)\right. \nonumber \\
& & \left. -x\sqrt{\frac{t}{\pi D}}e^{-x^2/4Dt}\right] .
	\end{eqnarray}

In our model we assume that the enamel is homogeneous at the initial moment. This assumption is not always fulfilled in real system. The initial concentrations are different for different tooth which is caused by individual features of each tooth such as hygiene, environment, etc. The experimental profiles corresponding to the first observation are also different from our initial condition (\ref{7}) in the cases considered in our paper. Taking into consideration the initial condition given by a specific function different from (\ref{7}) leads to the complications in the calculations and this can cause the perturbation method to be practically useless. Therefore, we assume that the initial condition can be chosen in the form given by Eq.~(\ref{7}), but we introduce a parameter $t_{\rm f}$ which is interpreted as a time interval during which the demineralization process leads form homogeneous concentration of {\it HA} to the `initial concentration' presenting in the plots. Of course, the parameter $t_{\rm f}$ is not a `real time' of demineralization process for a tooth situated in its natural environment; it represents a time of achieving the initial concentration in the experimental system. Thus, the time of observation is given as 
	\begin{displaymath}
t=t_{\rm f}+n\Delta t ,
	\end{displaymath}
where $n=0,1,2,\ldots$, and $\Delta t$ is the time step of successive concentration measurements. 

In Figs.~\ref{fig:fig5}--\ref{fig:fig6} the separated points represent the experimental data, the continuous lines are the approximate solutions given by Eq.~(\ref{stwym}) and the dashed lines are the numerical solutions of Eqs. (\ref{4})--(\ref{6}). The fitting parameters which ensure the best matching of approximate solutions (\ref{stwym}) to experimental data are $t_{\rm f}$, $D$ and the product $A_0k$. The most natural choice of the parameter $t_{\rm s}$ is $t_{\rm s}=1/(A_0k)$ what leads to $\varepsilon r=1$ [see Eqs.~(\ref{podst}) and~(\ref{15})] and $x_s=\sqrt{D/(A_0k)}$. In the papers \cite{daeg,gea} the initial concentration of $A$ is given in the units of ${\rm mol/cm^3}$ and there is rather impossible to express $A_0$ in the units used in the plots. Since the parameters $A_0$ and $k$ are not known, to obtain the numerical solutions we use the equations written in the dimensionless form (\ref{10})--(\ref{12}) with $\varepsilon r=1$. The values of $\varepsilon$ and $r$ are not determined unambiguously by experimental data. The numerical calculations performed for different values of the parameters show that the functions $c(\rho,\tau)$ obtained for any values of $\varepsilon$ under condition that $\varepsilon\leq 10^{-3}$ and $r=1/\varepsilon$ do not differ significantly from each other. The reason is that the time evolution of function $c(\rho,\tau)$ depends explicitly on the product $\varepsilon r$ [Eq. (\ref{12})] and its dependence on $\varepsilon$ is only manifested by the function $b(\rho,\tau)$, which for small $\varepsilon$ is very close to the function $b_0(\rho,\tau)$. So, we chose the parameters $\varepsilon=1/r=10^{-3}$ for numerical calculations. 
In considered cases the subsurface layer is rather thin in both of them. The height of this layer reaches about $30$ per cent of the initial values of the concentration of $HA$ at the second measurement and it decreases within time. So, we conclude that its influence on the concentration profiles (especially for long times) is small. For considered cases we assume that the width of the layer is very small comparing to the subsurface layer and we take $\rho_d=2$ to calculations. 
After finding the solutions of dimensionless equations, we transform them to the dimension form according to the formula (\ref{c}).

In Fig.~\ref{fig:fig5} the functions were calculated for the parameters $C_0=87\;{\rm mg/cm^2}$ and time interval $\Delta t=48\;{\rm h}$ taken from \cite{gea}. The fit parameters are found as $A_0k=1.5\times10^{-4}\;{\rm 1/s}$, $D=2.2\times10^{-10}\;{\rm cm^2/s}$ and $t_{\rm f}=5.5\;{\rm h}$. 
In Fig.~\ref{fig:fig6} we obtained the functions for $C_0=0.116\;{\rm g/cm^2}$, $D=9.9\times10^{-13}\;{\rm cm^2/s}$, $A_0k=1.5\times10^{-3}\;{\rm 1/s}$ and $t_{\rm f}=17\;{\rm h}$. 
As far as we know the exact values of the parameters $k$ and $D$ are unknown for the substances under considerations. Moreover, it is clear that these values depend on individual features of a tooth. Thus, we can not conclusively compare the obtained value of $D$ with the experimental one mentioned in the literature. However, we note that the parameter takes `typical' values. For example it was reported in \cite{dbd} that  $D\propto10^{-8}\div10^{-13}\;{\rm cm^2/s}$ for several substances diffusing in the tooth enamel.

We conclude this section by saying that the model qualitatively describes the data. The quantitative comparison is not fully conclusive but the fit is quite satisfactory.

\section{Final remarks}

We propose a theoretical model of the carious lesion progress, which is based on physically well-motivated diffusion-reaction equations. In contrast to the previous theoretical models of caries, the reaction term depends on the concentrations of diffusing acid and of enamel mineral. To approximately solve the diffusion-reaction equations we adopt a perturbation method. The concentration profiles presented in our paper are in qualitative agreement with the experimental data \cite{ac,itra,bl,cfh,ale,ae,abqde,ga,mzlkm,zkm,kpssm,tddg,gea,daeg} except in the region where the concentration of hydroxyapatite is the smallest. In the theoretical model the concentration of hydroxyapatite goes to zero in the region located just beyond the surface layer. In a real system (schematically shown in Fig.~\ref{fig:fig1}) a non-zero concentration of enamel is always observed. This is caused by occurrence of various impurities, such as fluoride, in the enamel. These impurities prevent a complete loss of the enamel. However, taking the impurities into considerations significantly complicates the model. We are mostly interested in modeling of caries in the region, where changes of {\it HA} concentration are the largest because therein the caries progress is most intensive and the limit of caries is located in this region. Thus, we do not take into account the enamel impurities in the considered model. Analyzing the experimental concentration profiles of {\it HA} presenting in the paper cited above, one can see that the surface layer is often created in the form differing from the one assumed in our model, for example the {\it HA} concentration decrease in time in this layer. Moreover, the initial concentration of {\it HA} is not given by Eq. (\ref{7}). In our opinion the model considered in our paper can be used to describe the caries in such situations. We observed that the reduction of the {\it HA} concentration in surface layer does not noticeably change the solutions of the diffusion--reaction equations. Introduction of the parameter $t_{\rm s}$ allows one to find the solution of the equations corresponding to the initial concentration in the experiment.

A quantitative comparison requires returning to the dimensional variables $x$ and $t$. This is possible if one knows values of the parameters occurring in Eqs. (\ref{4})-(\ref{6}). As far as we know, the values of these parameters are unknown for teeth studied in the papers presenting the measured concentrations. Qualitative comparison is also difficult because of individual features of each tooth caused by various factors such as age, environment, hygiene, etc., which result in the different quantitative characteristics of each tooth. Thus, repeatability of the experimental results is weak. The changes of concentration of the acid are controlled mainly by the parameter $\varepsilon$ only, whereas the loss of hydroxyapatite mostly depends on the product of the parameters $\varepsilon r$. We note that the parameter $r=A_0/C_0$ is controlled in experiments {\it in vitro} because $A_0$ is the acid concentration of the buffer solution. 

The perturbation method was already used to solve the diffusion--reaction equations for the homogeneous system consisting of two mobile initially separated reactants~\cite{thktw,tkhkw,tykhw}, where, similarly to our paper, up to the first order solutions were obtained. However, the first order solutions in \cite{thktw,tkhkw,tykhw} were calculated only within long time approximation. Let us note that in our case of one static and one mobile reactant, we found the exact solutions up to the first order within the perturbation method.

\begin{acknowledgments}
The authors wish to express their thanks to Stanis{\l}aw Mr\'owczy\'nski for fruitful discussions and critical comments on the manuscript. This work was supported in part by Grant 1 P03B 136 30 from the Polish Ministry of Education and Science.
\end{acknowledgments}

\bibliography{lewandowska}

\end{document}